\documentclass{ws-ijmpa}

\def\Cite#1{\protect\cite{#1}}
\def\etal{{\sl et al.}}
\def\lnA{\langle\ln A\rangle}
\def\Xmax{$X_{max}$}
\def\fref#1{Fig.~\ref{#1}}
\def\dotted{$\cdot\cdot\cdot$}
\def\dashed{-\,-\,-}
\def\line{---}
\def\lleft{{\sl left}}
\def\rright{{\sl right}}
\def\LLeft{{\sl Left}}
\def\RRight{{\sl Right}}

\begin{document}

\markboth{J.R. H\"orandel}{Overview on direct and indirect measurements}

\catchline{}{}{}{}{}

\title{Overview on direct and indirect measurements of cosmic rays\\
       \textit{Some thoughts on galactic cosmic rays and the knee}
       \footnote{Invited overview, presented at the 19th European
       Cosmic Ray Symposium, August 30$^{th}$ - September 3$^{rd}$, 2004,
       Florence, Italy.}
       }

\author{J\"org R. H\"orandel}
\address{University of Karlsruhe, Institut f\"ur Experimentelle Kernphysik,
         PO 3640, 76021 Karlsruhe, Germany, hoerandel@ik.fzk.de}

\maketitle


\pub{Received (Day Month Year)}{Revised (Day Month Year)}

\begin{abstract}
 An overview is given on results from direct and indirect measurements of
 galactic cosmic rays. Their implications on the contemporary understanding of
 the origin of cosmic rays and the knee in their energy spectrum are discussed.
\end{abstract}


\section{INTRODUCTION}
Cosmic rays (CRs) in the energy range from several GeV up to about 100~PeV are
assumed to be mostly of galactic origin.  At energies up to several 100~MeV
individual isotopes can be identified, e.g.  with the ACE/CRIS
experiment\cite{cris}, a satellite borne silicon detector telescope.  At higher
energies, CRs are identified by their charge and the energy measurement becomes
an experimental challenge.  Various techniques are utilized, like the
determination of the particles momenta in magnet spectrometers (e.g.
BESS\cite{bess}), the (partial) absorption of nuclei in calorimeters (e.g.
ATIC\cite{atic}), or the measurement of transition radiation emitted by
relativistic particles (e.g. TRACER\cite{gahbauer}).  Circumpolar long duration
balloon flights offer the possibility of a long exposure ($\ge14$~d) combined
with low atmospheric overburden (typically $<5$~g/cm$^2$) as recently
demonstrated by the ATIC\cite{atic03}, TIGER\cite{tiger03}, and
TRACER\cite{jrhcospar} experiments.

At energies above 1~PeV the steeply falling spectrum requires large detection
areas (exceeding several $10^4$~m$^2$) and exposure times of several years,
which presently can be realized only in ground based installations.  They
measure the secondary products generated by the CR particles in the atmosphere
-- the extensive air showers.  The challenge of these investigations is to
reveal the properties of the primary particle behind an absorber -- the
atmosphere -- with a total thickness, corresponding to 11 hadronic interaction
lengths or 30 radiation lengths.  Consequently, these experiments have a
coarser resolution and only mass groups or the average primary mass are
derived.  Two basic approaches can be distinguished: Measuring the debris of
the particle cascade at ground level by registering the main shower components,
the electromagnetic, muonic, and hadronic parts. Or measuring the longitudinal
shower development in the atmosphere, exploring the \v{C}erenkov or
fluorescence light generated predominantly by the shower electrons.  Examples
are the KASCADE\cite{kascadenim}-Grande\cite{grande} or EAS-TOP\cite{eastop}
installations, measuring simultaneously the electromagnetic, muonic, and
hadronic shower components, the SPASE\cite{spase}/AMANDA\cite{amanda}
experiment, investigating electrons and high energy muons, or the
BLANCA\cite{blanca} \v{C}erenkov detector.

In this article an overview on results from direct and indirect measurements is
given, concerning the sources of CRs (\S\ref{sourcesect}), their propagation
through the Galaxy (\S\ref{propsect}), and the energy spectra and mass
composition observed at Earth (\S\ref{espeksect}). Their implications on the
understanding of the origin of the knee are discussed (\S\ref{kneesect}).

\section{SOURCES OF COSMIC RAYS} \label{sourcesect}

A big step towards the understanding of CR sources would be their direct
observation in the sky. However, charged CRs are deflected in the galactic
magnetic fields, the gyromagnetic radius of a proton with an energy of 1~PeV is
about 0.4~pc. But $\gamma$-rays, are good candidates for a point source search.
Photon emission of supernova remnants (SNRs) has been detected in a wide energy
range from radio wave lengths to x-rays.  The observations are interpreted as
synchrotron emission from electrons, which are accelerated in these
regions\cite{berezhko-casa}. The HEGRA experiment\cite{hegra-casa} has detected
an excess of $\gamma$-rays with TeV energies from the SNR Cassiopeia~A.  This
is interpreted as evidence for hadron acceleration in the SNR. The hadrons
interact with protons of the interstellar medium close to the source region,
producing $\pi^0$s, which decay into high-energy photons.  The flux is
compatible with a model of electron and hadron acceleration in shock fronts of
the SNR\cite{berezhko-casa}.

\begin{figure}
  \psfig{file=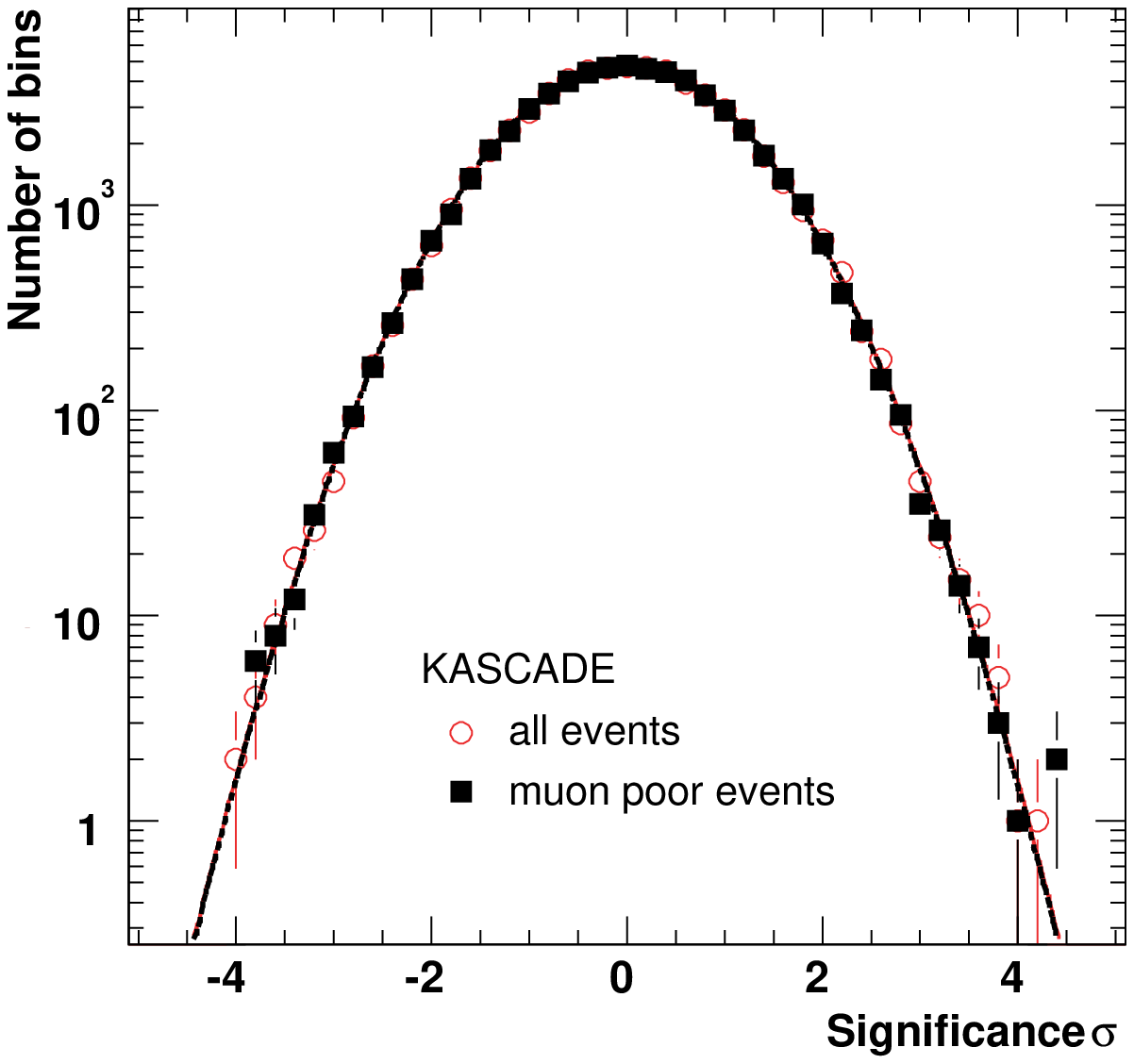,width=0.51\textwidth}
  \psfig{file=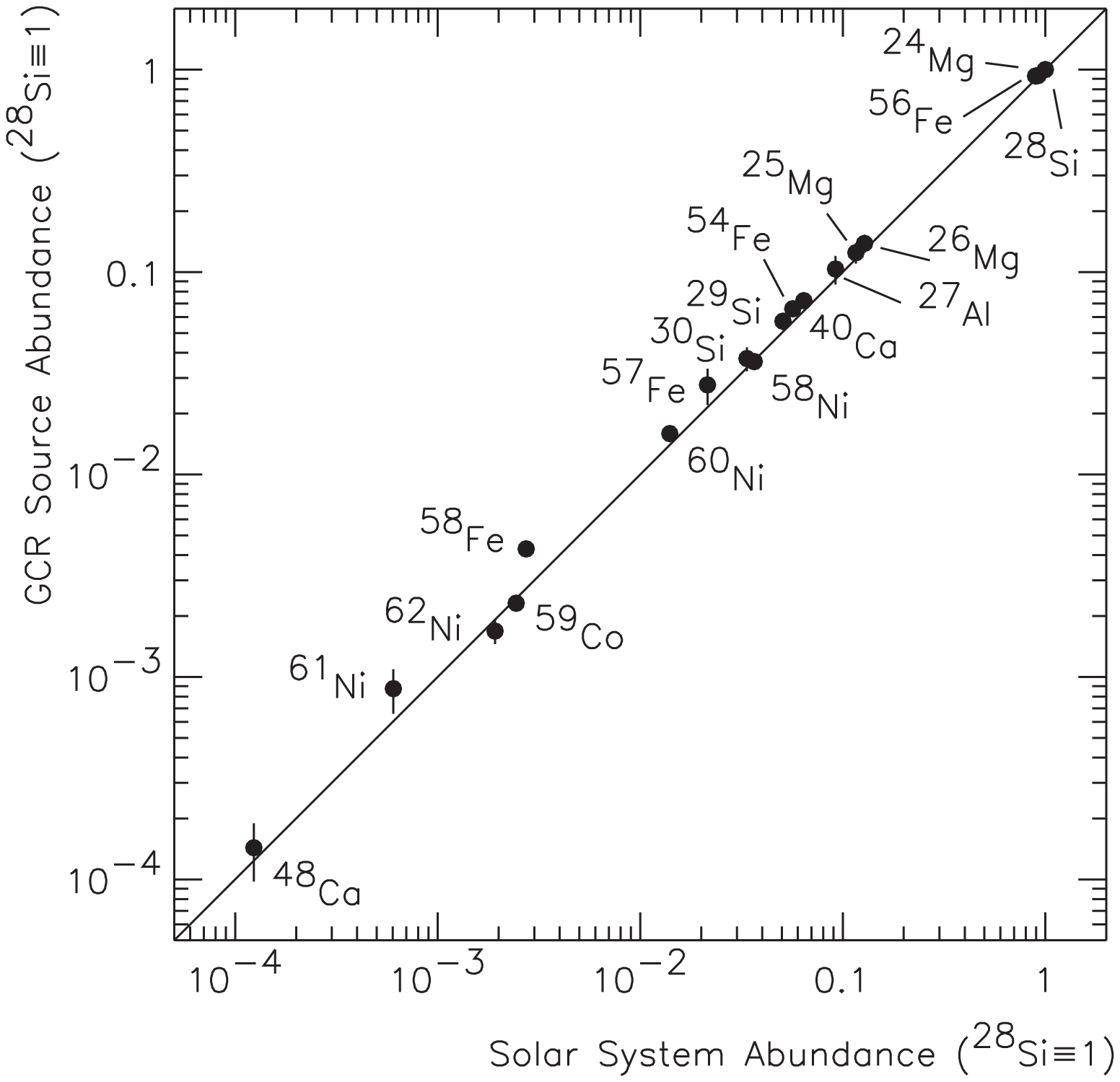,width=0.47\textwidth}
  \caption{\LLeft:  Distribution of the significance values from a sky map of
	   the arrival direction of CRs as measured by the KASCADE
	   experiment\Cite{kascade-points} for the complete data set (open
	   circles) and a selection of muon poor showers (filled squares).
  \RRight: Comparison of derived CR source abundances of refractory
	   nuclides with solar-system abundances according to measurements
	   with ACE/CRIS\Cite{cris-abundance} normalized to $^{28}$Si.}
 \label{points}
\end{figure}

Despite of the above-mentioned deflection, it is of great interest to study the
arrival direction of charged CRs as well.  The result of such an analysis from
the KASCADE experiment\cite{kascade-points} is depicted in \fref{points}
(\lleft).  Shown is the distribution of the significances from a sky map of the
arrival direction of showers with energies above 0.3~PeV covering a region from
$10^\circ$ to $80^\circ$ in declination.  For an isotropic distribution the
significances are expected to follow a Gaussian distribution as indicated by
the solid line.  Results for all events are presented, as well as for a
selection of muon-poor showers. The latter are expected from potential primary
$\gamma$-rays. No significant deviation of the data from the Gaussian
distribution can be recognized.  The analysis has been deepened by
investigating a narrow band ($\pm1.5^\circ$) around the galactic plane. Also
circular regions around SNRs and TeV-$\gamma$-ray sources have been
studied.  None of the searches provided a hint for a point source.  In
addition, no clustering of the arrival direction for showers with primary
energies above 80~PeV is visible.
Claims by the MAKET-ANI experiment for a point-source
detection\cite{maketani-points} have been withdrawn
meanwhile\cite{chiliconter}.

Despite no sources have been detected with charged particles, information on
the composition at the source can be obtained from measurements of the
abundance of refractory nuclei. They appear to have undergone minimal elemental
fractionation relative to one another.  The derived abundance at the source is
presented in \fref{points} (\rright) versus the abundance in the solar
system\cite{cris-abundance}.  The two samples exhibit an extreme similarity
over a wide range. Of the 18 nuclides included in this comparison, only
$^{58}$Fe is found to have an abundance relative to $^{28}$Si that differs by
more than a factor of 1.5 from the solar-system value. When uncertainties are
taken into account, all of the other abundances are consistent with being
within 20\% of the solar values. This indicates that CRs are accelerated out of
a sample of well mixed interstellar matter.

Motivated by the observations, it is assumed that at least a large fraction of
CRs are accelerated in supernova
remnants\cite{berezhko,stanev,kobayakawa,sveshnikova}.  However, recent
progress in the understanding of $\gamma$-ray bursts has put forward the idea
that a subsample of high-energy CRs may be accelerated in $\gamma$-ray
bursts\cite{wick,dar}.

\section{PROPAGATION OF COSMIC RAYS} \label{propsect}
After acceleration, the particles propagate in a diffusive process through the
Galaxy, being deflected many times by the randomly oriented magnetic
fields ($B\sim3$~$\mu$G). 
The nuclei are not confined to the galactic disc, they propagate in the
galactic halo as well. The diffuse $\gamma$-ray background, extending well
above the disc, detected by the EGRET experiment, exhibits a structure in the
GeV region, which is interpreted as indication for the interaction of
propagating CRs with interstellar matter\cite{strong-moskalenko}.
The height of the halo has been estimated with measurements of the
$^{10}$Be/$^9$Be-ratio by the ISOMAX experiment\cite{simon-height} to be a few
kpc.
The measured abundance of radioactive nuclei in CRs with the CRIS
instrument yields a residence time in the Galaxy of about $15\cdot10^6$~a for
particles with GeV energies\cite{cris-time}. 

\begin{figure}
  \psfig{file=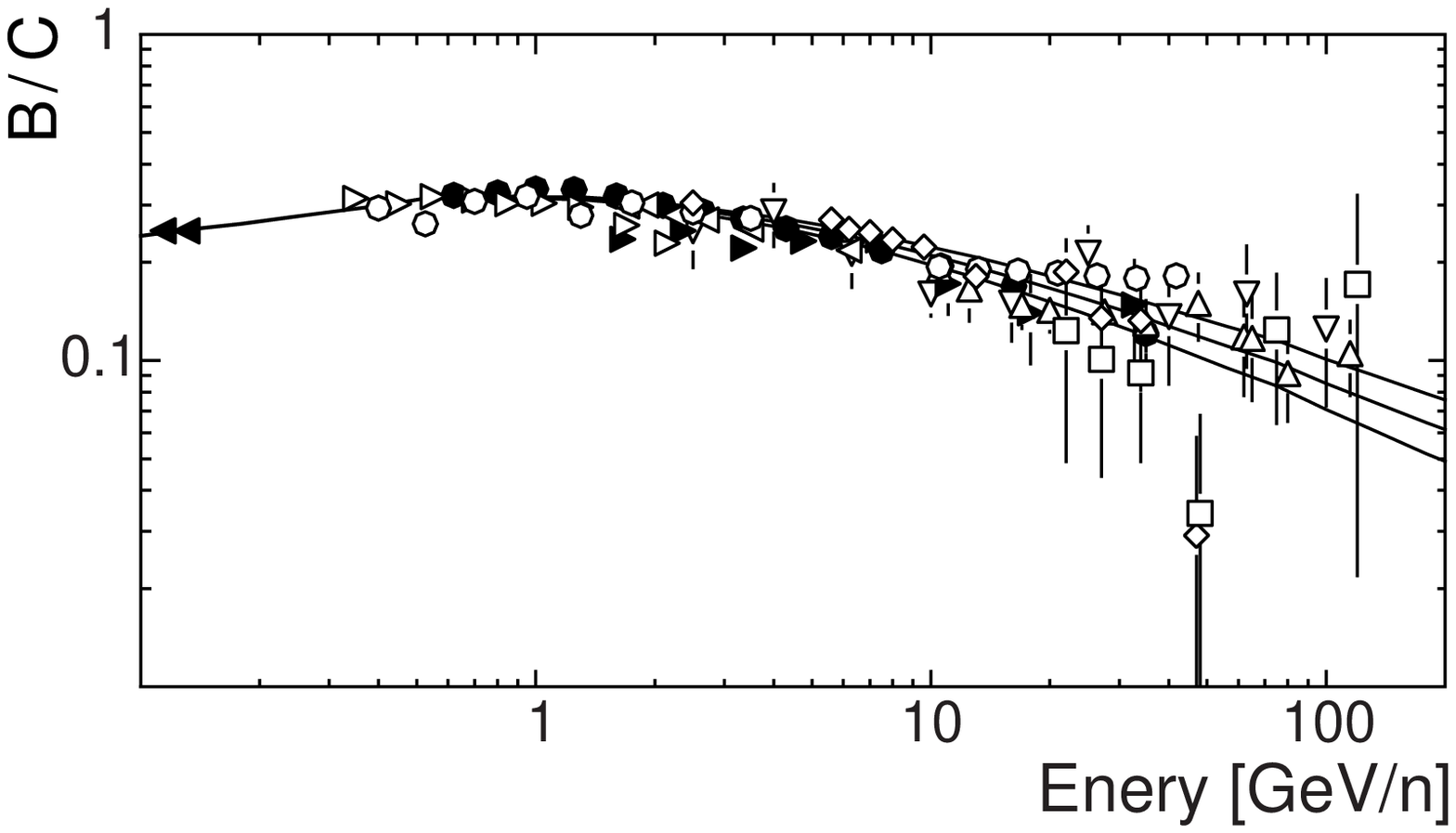,width=0.49\textwidth}
  \psfig{file=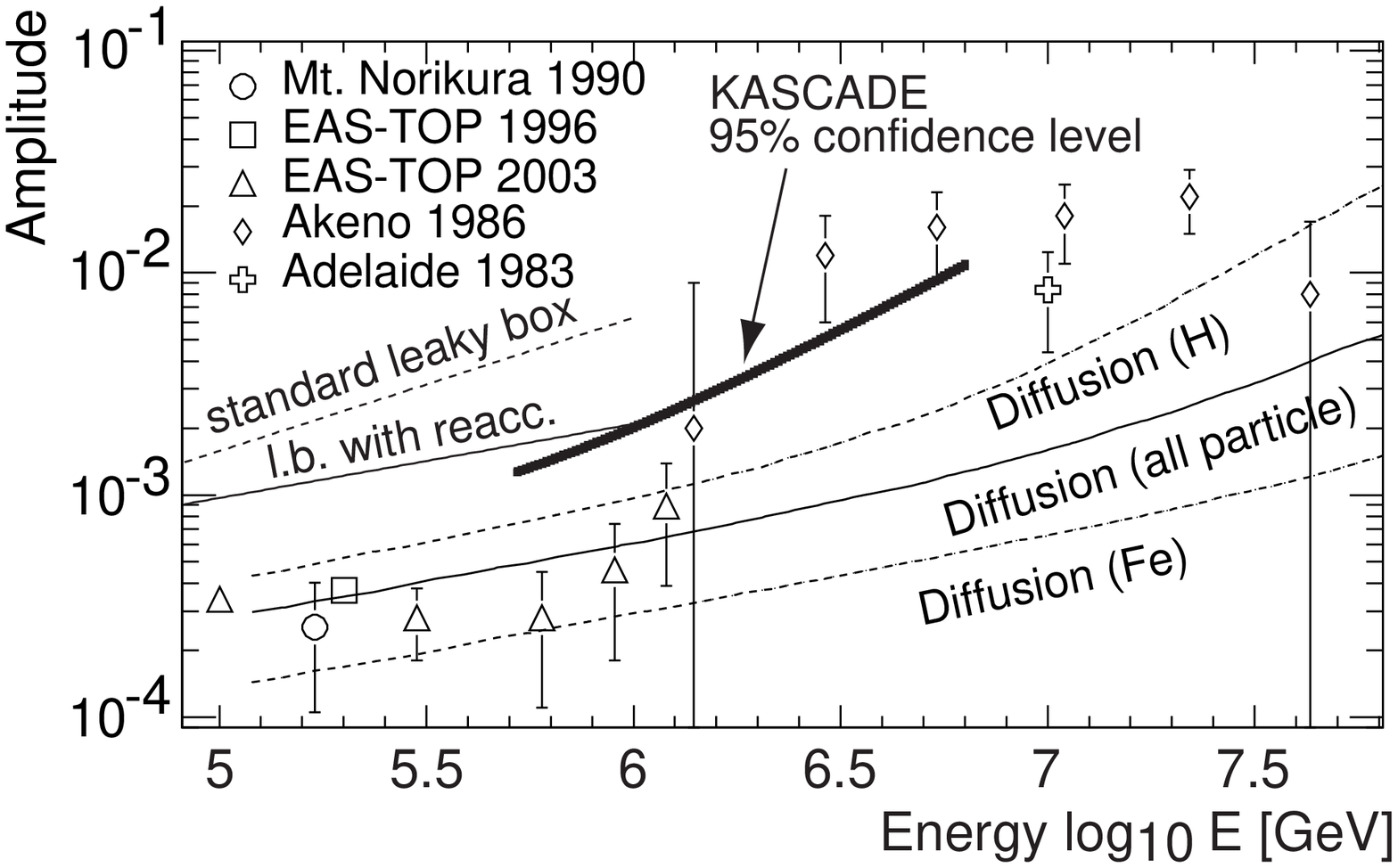,width=0.49\textwidth}
  \caption{\LLeft: Measured boron-to-carbon ratio as function of energy,
	   the lines indicate model predictions, see\Cite{stephens}.  
	   \RRight: Rayleigh amplitudes as function of energy for various
	   experiments, for references see\Cite{kascade-aniso}. Additionally,
	   model predictions for Leaky Box models\Cite{ptuskinaniso} and a
	   diffusion model\Cite{candiaaniso} are shown. The lines indicate the
	   expected anisotropy for primary protons, iron nuclei, and all
	   particles.}
 \label{aniso}
\end{figure}

Information on the propagation pathlength of CRs is often derived from the
measurement of the ratio of primary to secondary nuclei. The latter are
produced through spallation during propagation in the Galaxy.  As an example,
the measured boron-to-carbon ratio is shown in \fref{aniso} (\lleft) as
function of energy\cite{stephens}. The energy dependence of the measured
ratio is frequently explained in Leaky Box models by a decrease of the
pathlength of CRs in the Galaxy $\Lambda(R) = \Lambda_0 (R/R_0)^{-\delta}$,
with typical values $\Lambda_0\approx10 - 15$~g/cm$^2$, $\delta\approx0.5 -
0.6$, and the rigidity $R_0\approx4$~GV.

At higher energies such measurements are not feasible due to the limited mass
resolution of air shower experiments.  However, at these energies the large
scale anisotropy is expected to reveal properties of the CR propagation.  The
Rayleigh formalism is applied to the right ascension distribution of extensive
air showers measured by KASCADE\cite{kascade-aniso}.  No hints of anisotropy
are visible in the right ascension distributions in the energy range from 0.7
to 6~PeV. This accounts for all showers, as well as for subsets containing
showers induced by predominantly light or heavy primary nuclei. Upper 
limits are shown together with results from other experiments in \fref{aniso}
(\rright). It presents the Rayleigh amplitude as function of energy.  The
experimental results are compared to the anisotropy expected from calculations
of the propagation of CRs in the Galaxy. 
The data reflect a trend predicted by a diffusion
model\cite{candiaaniso}.  This indicates that leakage from the Galaxy and
consequently a decreasing pathlength $\Lambda(E)$ plays an important part
during CR propagation at high energies and most likely, also for the origin of
the knee.

Leaky Box models are successful at GeV energies as discussed above.  In the PeV
regime, however, they seem to be faced with some difficulties.  Two versions of
a Leaky Box model\cite{ptuskin}, with and without reacceleration, seem to be
ruled out by the anisotropy measurements, see \fref{aniso}. This relates to the
extremely steep decrease of the pathlength $\Lambda\propto E^{-0.6}$, yielding
at PeV energies unrealisticly small values for $\Lambda$. Even for a residual
pathlength model\cite{swordy}, at 1~PeV the pathlength would be smaller than
the matter traversed along a straight line from the center of the Galaxy to the
solar system\cite{jrhprop}.

\section{ENERGY SPECTRA AND MASS COMPOSITION} \label{espeksect}

\begin{figure}
  \psfig{file=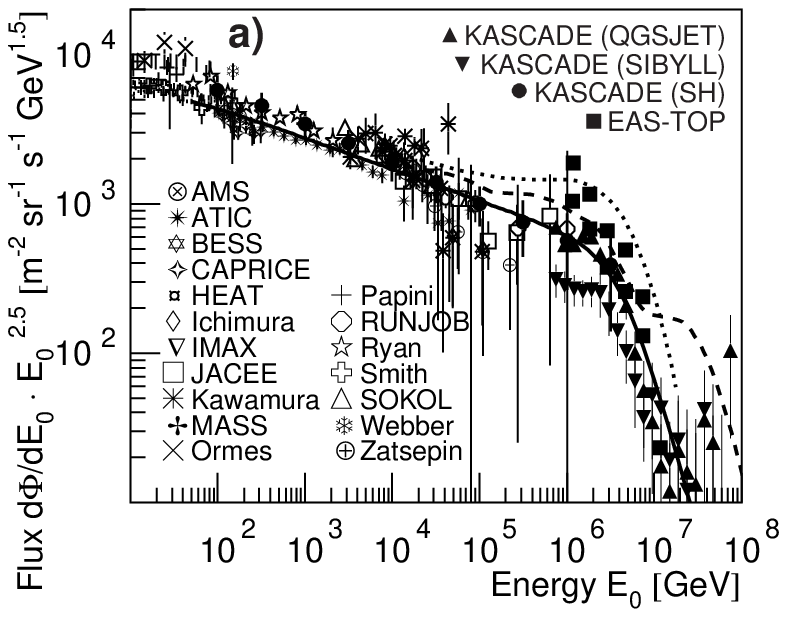,width=0.49\textwidth}
  \psfig{file=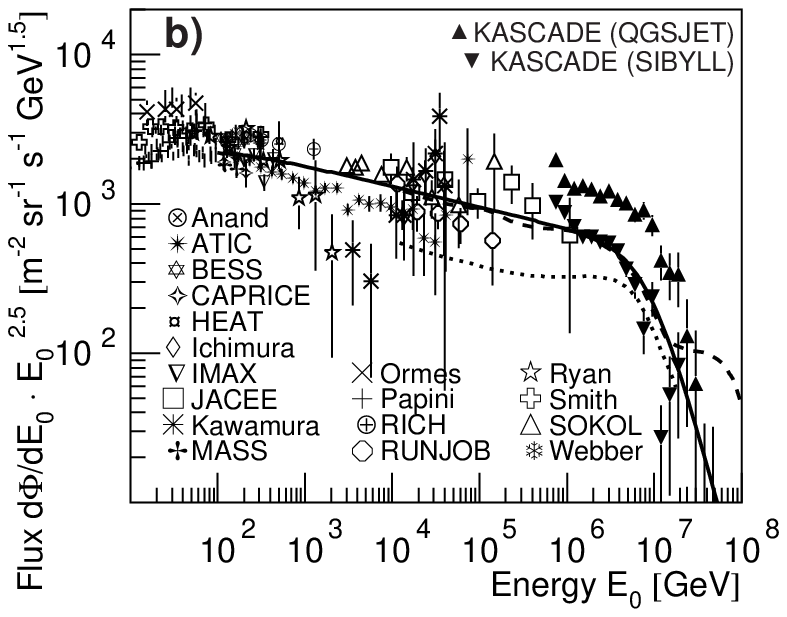,width=0.49\textwidth}
  \psfig{file=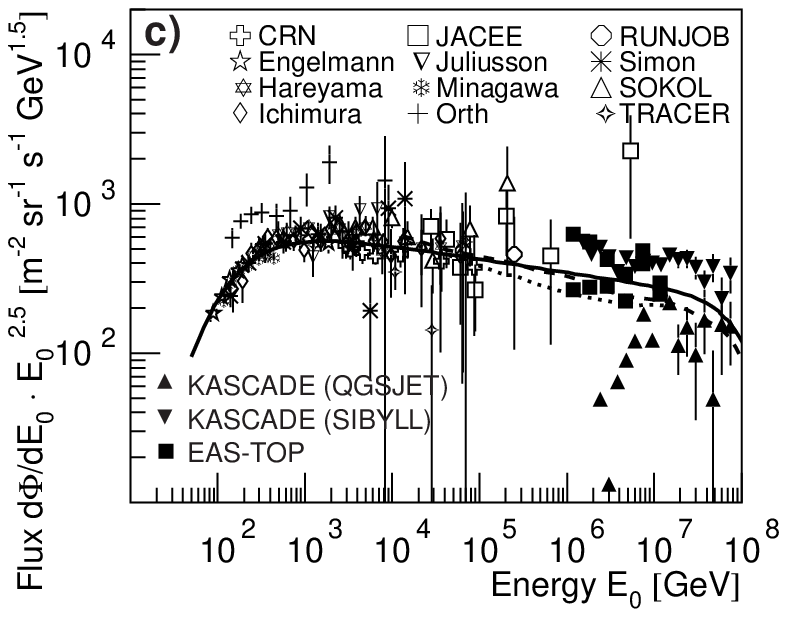,width=0.49\textwidth}
 \begin{minipage}[b]{0.49\columnwidth}
 \caption{Energy spectra for elemental groups {\bf a)}
	  protons, {\bf b)} helium, and {\bf c)} iron.  Open symbols give
	  results of direct measurements, for references
	  see\Cite{pg,gahbauer,atic03}.  Filled
	  symbols represent data from air shower measurements: KASCADE
	  electrons/muons interpreted with two interaction
	  models\Cite{ulrichepj} (preliminary), KASCADE single
	  hadrons\Cite{kascadesh}, and EAS-Top
	  electrons/muons\Cite{eastop-knee}. The data are compared to
	  calculations by Kalmykov \etal\Cite{kalmykov} (\dotted),
	  Sveshnikova\Cite{sveshnikova} (\dashed), and the Poly-Gonato
	  model\Cite{pg} (\line).}
 \label{elementspek}
 \end{minipage}
\end{figure}

At energies below 100~TeV the energy spectra of individual elements have been
measured with detectors above the atmosphere. Examples for protons, helium and
iron nuclei are compiled in \fref{elementspek}. The measured spectra can be
described by powerlaws.  For the iron spectrum at low energies the modulation
due to the magnetic fields of the heliosphere causes the flux suppression.
Actual experiments, like ATIC\cite{atic03} and TRACER\cite{jrhcospar}, as well
as the proposed ACCESS\cite{access} space project are expected to improve the
experimental situation in the region around 0.1 to 1~PeV, where large
uncertainties are visible in the figure.  More precise fluxes in this region
would be valuable to intercalibrate air shower measurements. 

\begin{figure}
 \psfig{file=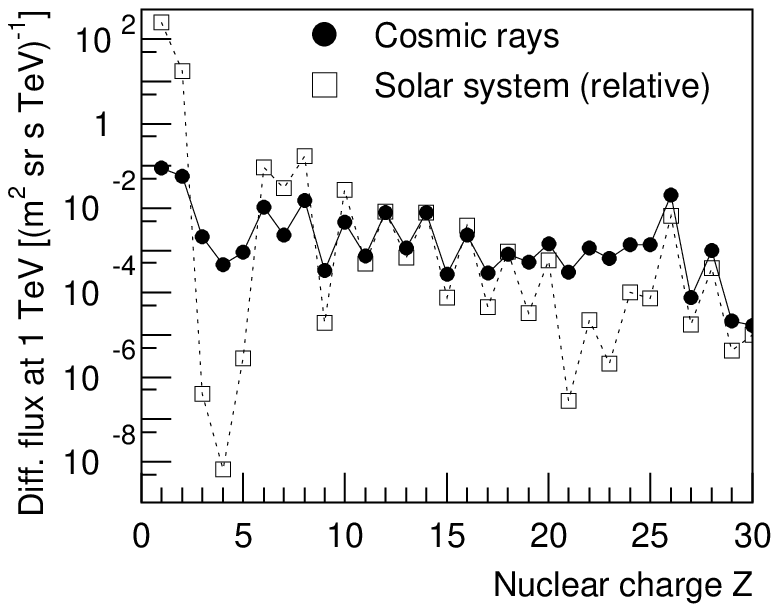,width=0.49\textwidth}
 \psfig{file=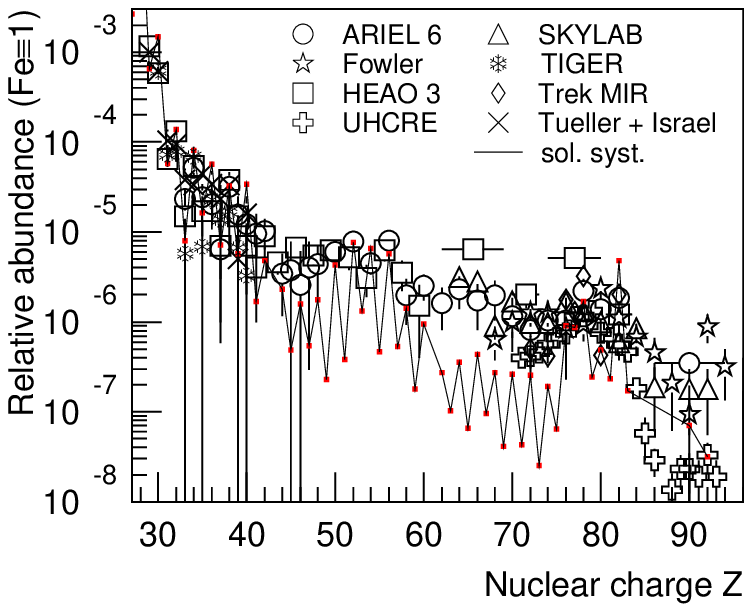,width=0.49\textwidth}
 \caption{\LLeft: Abundance of elements ($Z\le28$) in CRs\Cite{wiebel,pg} at 
	  1~TeV.
	  \RRight: Relative abundance of CR elements ($Z>28$) normalized to
	  Fe$\equiv 1$ from various experiments around 1~GeV/n.  For references
	  see\Cite{pg,tiger03}.  For comparison, abundances in the solar
	  system\Cite{cameron} are presented as well, normalized to Si (\lleft)
	  and to Fe (\rright).} 
 \label{abundance}
\end{figure}

The elemental abundance at 1~TeV is presented in \fref{abundance} (\lleft) as
function of nuclear charge number for elements up to nickel.  The experimental
status for the heavier elements is summarized in \fref{abundance} (\rright).
All stable elements of the periodic table have been registered in CRs.  In both
panels the CR abundance is compared to the abundance in the solar
system\cite{cameron} normalized to silicon and iron, respectively. The overall
similarity of the two samples of matter, already seen in \fref{points}, is
reflected here on a coarser scale.

\begin{figure}
  \psfig{file=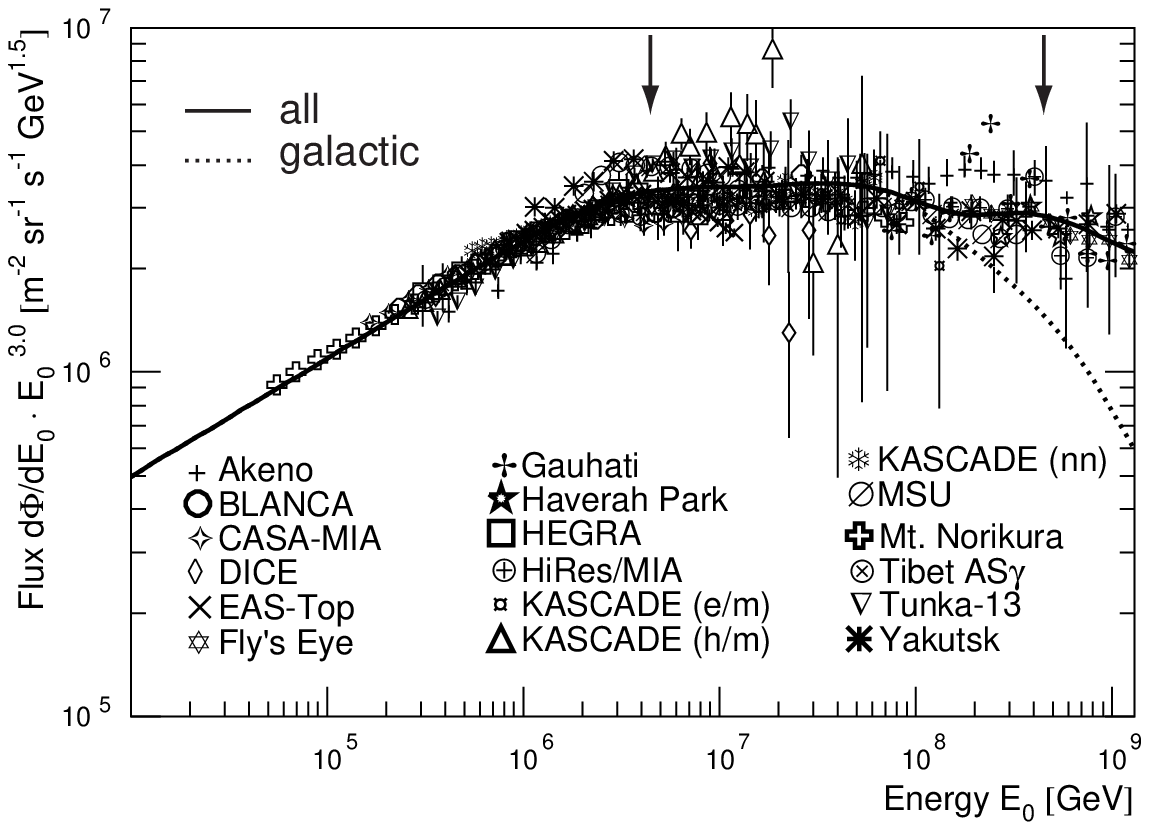,width=0.50\textwidth}
  \psfig{file=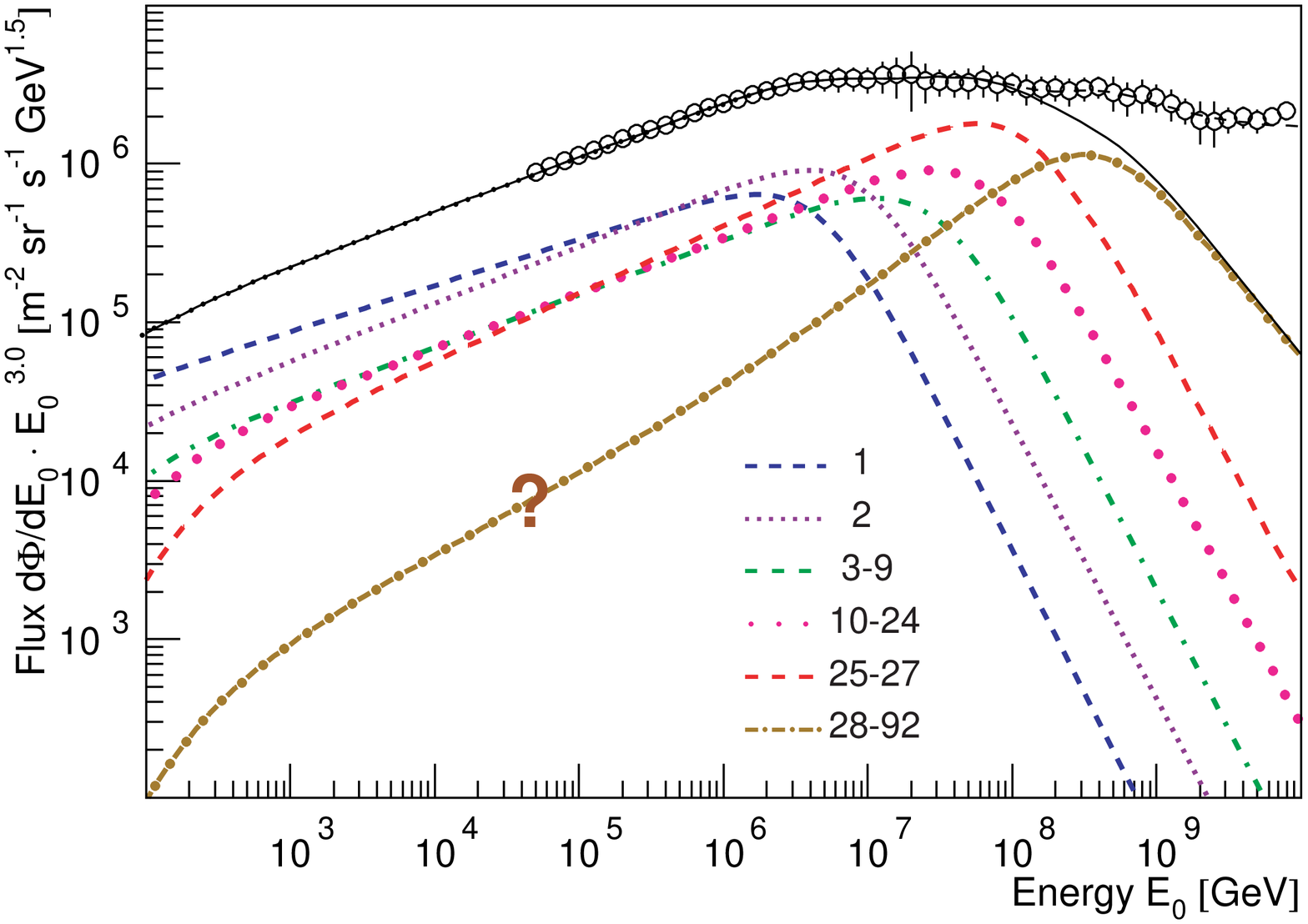,width=0.48\textwidth}
  \caption{\LLeft: Normalized all-particle energy spectra from different
	   experiments. The lines indicate the average all particle spectrum
	   and the contribution of galactic CRs.  The knee at
	   $E_k\sim4.5$~PeV and the second knee at $\sim400$~PeV$\approx92\cdot
	   E_k$ are indicated. 
	   \RRight: The average flux of the measurements (\lleft) is
	   represented by the data points.  Additionally, spectra for elemental
	   groups with the indicated charge number range according to a
	   parameterization of the measurements are depicted, including a
	   proposed contribution of ultra-heavy elements ($Z>28$),
	   extrapolated from measurements at GeV energies ("?").
	   For details and references see\Cite{pg}.} 
 \label{knie}
\end{figure}

At higher energies, many air shower experiments have reported fluxes for all
particles.  A compilation is presented in \fref{knie} (\lleft). The energy
scale of the individual experiments has been slightly normalized ($\pm10\%$) in
order to match the flux with that obtained by direct measurements\cite{pg}.  A
good agreement between the experiments in the reconstructed shape of the
spectrum is evident.  The knee at $\sim4.5$~PeV and a smaller structure at
$\sim400$~PeV, the second knee, are visible.

Most valuable to reveal the origin of the knee are measurements of the energy
spectra for individual elements or at least elemental groups. KASCADE studied
the influence of different hadronic interaction models used in the simulations
to interpret the data\cite{ulrichepj}.  Two sets of spectra, derived from the
observation of the electromagnetic and muonic air shower components, applying
an unfolding procedure based on the Gold algorithm and using
CORSIKA\cite{corsika} with the hadronic interaction models QGSJET and SIBYLL
are compiled in \fref{elementspek} for three elemental groups.  As can be seen
in the figure, the fluxes depend on the model used.  KASCADE emphasizes that,
at present, there are systematic differences between measured and simulated
observables which cause the ambiguities of the spectra. These conclusions apply
in a similar way also to other experiments.  A correct deconvolution of energy
spectra requires a more precise knowledge of the hadronic interactions.  

\fref{elementspek} also shows the spectrum of primary protons, which has been
derived from the flux of unaccompanied hadrons measured by
KASCADE\cite{kascadesh}.  The spectrum is compatible with the proton flux as
obtained from the unfolding procedure when using the QGSJET model.  The EAS-TOP
experiment published two sets of spectra with different assumptions about the
contribution of protons and helium nuclei derived from the measurements of the
electromagnetic and muonic shower components\cite{eastop-knee}. The resulting
fluxes are indicated by two squares per primary energy.  To guide the eye, the
solid lines indicate power law spectra with a cut-off at $Z\cdot4.5$~PeV.

The dashed lines represent calculations of energy spectra for nuclei
accelerated in supernovae\cite{sveshnikova}.  It
is assumed that the particles are accelerated in a variety of supernovae
populations, each having an individual maximum energy that can be attained
during acceleration, which results in the bumpy structure of the obtained
spectra.  The dotted lines reflect calculations of the diffusive propagation of
particles through the Galaxy\cite{kalmykov}. The leakage of particles yields a
rigidity dependent cut-off.
Comparison with the data may suggest a {\sl qualitative} understanding of the
energy spectra. However, for a precise {\sl quantitative} understanding,
detailed investigations of the systematic errors of the measurements are
necessary and the description of the interaction processes in the atmosphere
needs to be improved. 

\begin{figure}
 \psfig{file=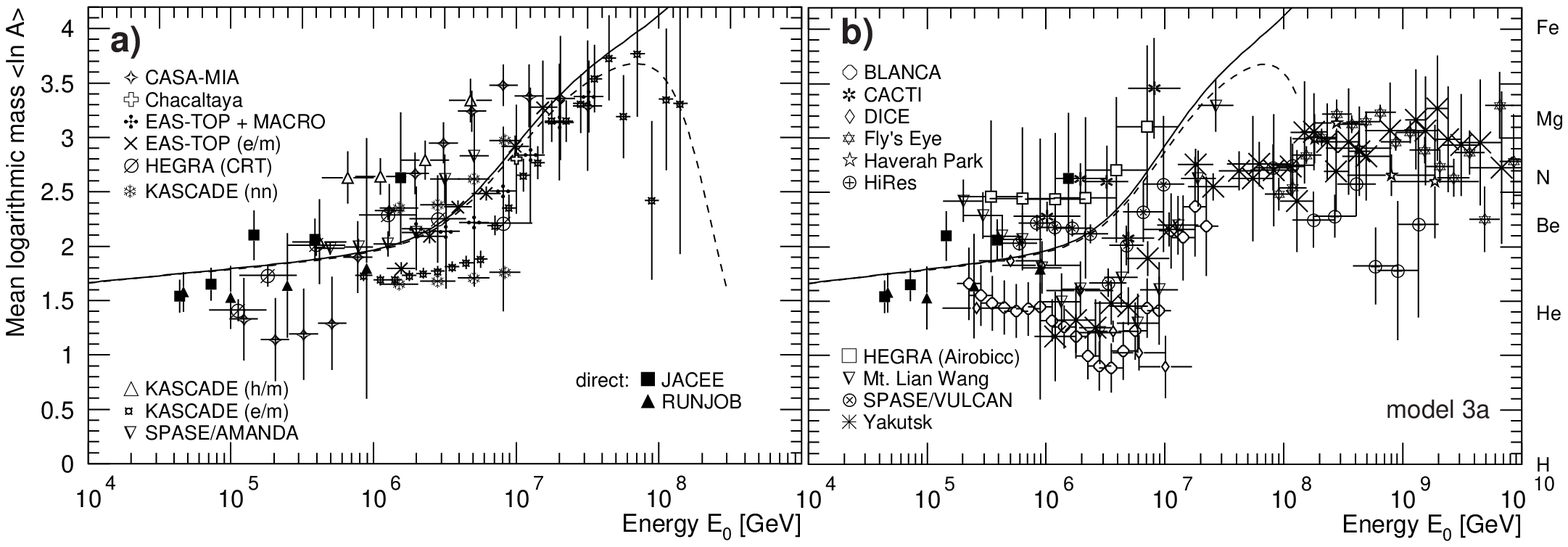,width=\textwidth}
 \caption{Mean logarithmic mass of CRs reconstructed from
	  {\bf a)} experiments measuring electrons, muons, and hadrons at
	  ground level\Cite{pg,eastop-em-lna,eastop-macro-lna,spaseamandalna}
	  and
	  {\bf b)} observations of the shower maximum interpreted with a
	  modification of QGSJET\Cite{wq}.  The lines indicate expectations
	  according to the Poly Gonato model\Cite{pg}.} 
 \label{masse}
\end{figure}

While the elemental abundance is relatively well known at low energies from
direct measurements (see \fref{abundance}), at higher energies, air-shower
experiments provide information on mass groups or on the average mass.
Frequently, the mean logarithmic mass $\lnA$, defined as $\lnA=\sum r_i\ln
A_i$, where $r_i$ is the relative fraction of nuclei with atomic mass number
$A_i$, is used to characterize the composition.  $\lnA$ is often derived from
the ratio of particles measured at ground level.  For a primary proton more
electrons and hadrons and fewer muons are registered as compared to an iron
induced shower with the same energy.  The data from many experiments are
compiled in \fref{masse} (\lleft). They exhibit an increase of $\lnA$ as
function of energy in the knee region.  The increase is compatible with
expectations, assuming a cut-off behavior of the flux of individual elements as
indicated in \fref{elementspek} by the solid line.  The second class of
experiments reconstructs the average depth of the shower maximum \Xmax\ from
the observation of \v{C}erenkov and fluorescence light. Using the model QGSJET
to derive the mean logarithmic mass from the data results in a light mass
composition at high energies in contradiction to the findings just
mentioned\cite{pg}. Introducing modifications to QGSJET, namely lowering the
inelastic cross sections and slightly increasing the elasticity of hadronic
interactions, this discrepancy can be reduced\cite{wq} and the mean logarithmic
mass rises as function of energy, see \fref{masse} (\rright).

\begin{figure}
  \psfig{file=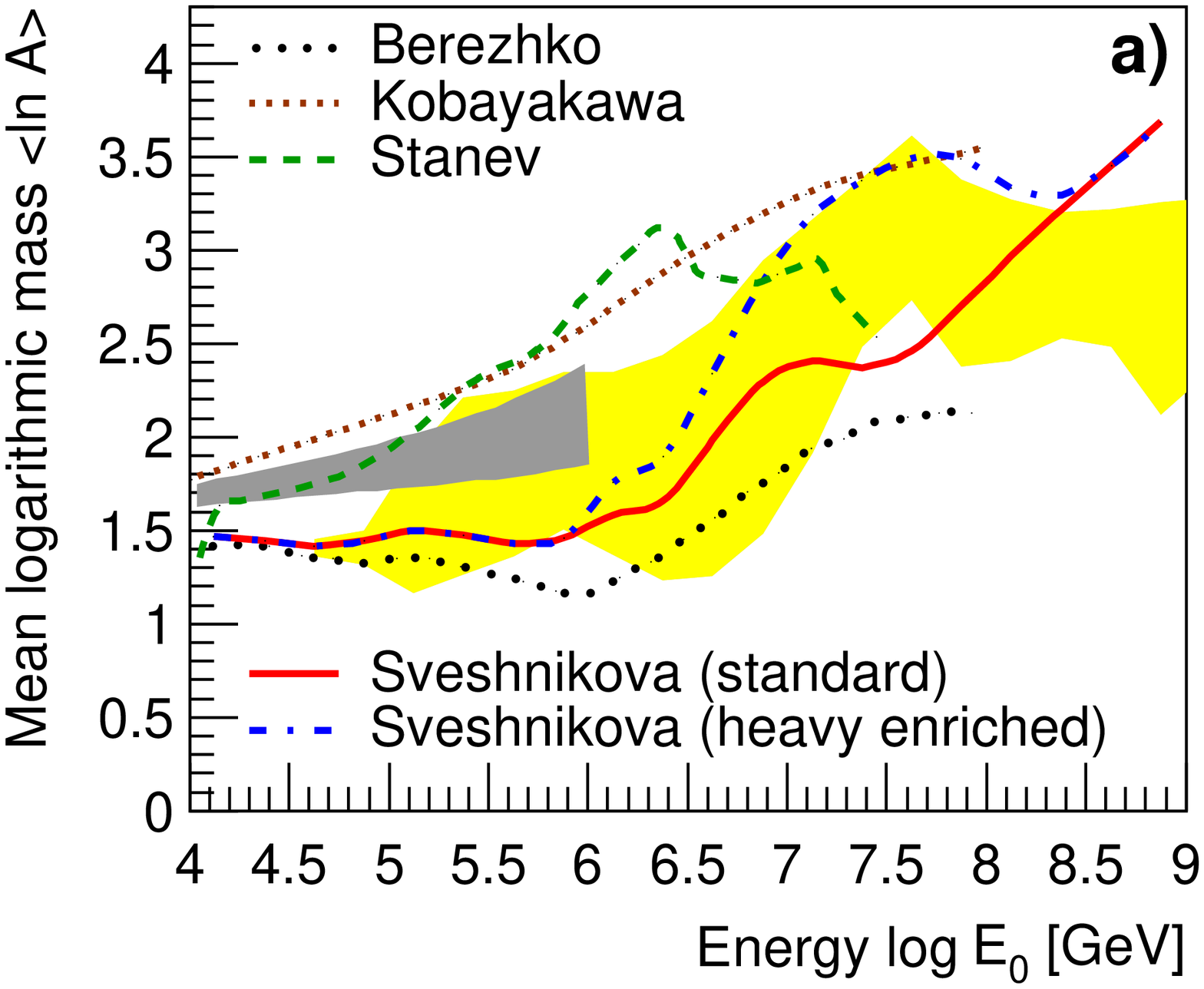,width=0.49\textwidth}
  \psfig{file=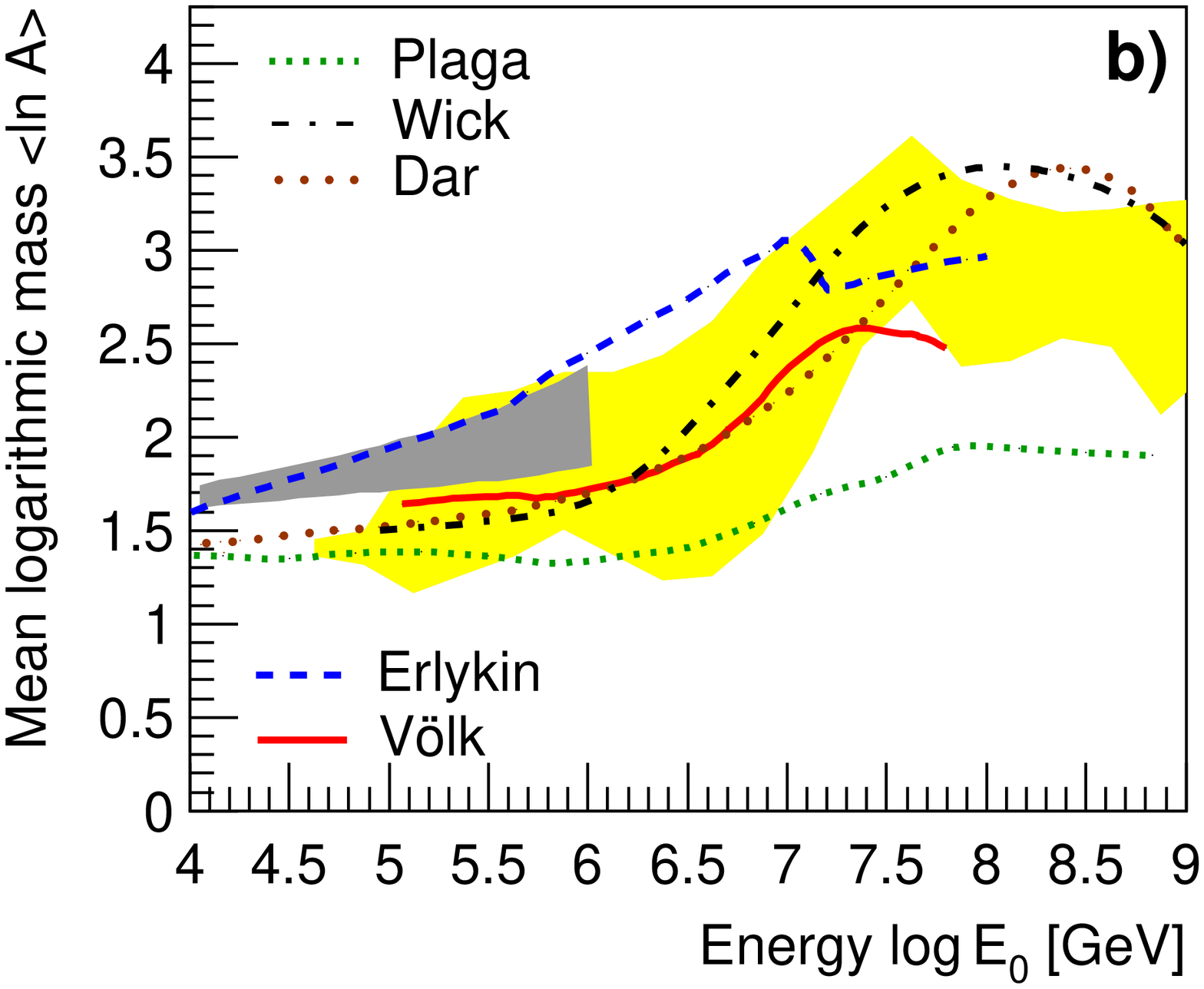,width=0.49\textwidth}
  \psfig{file=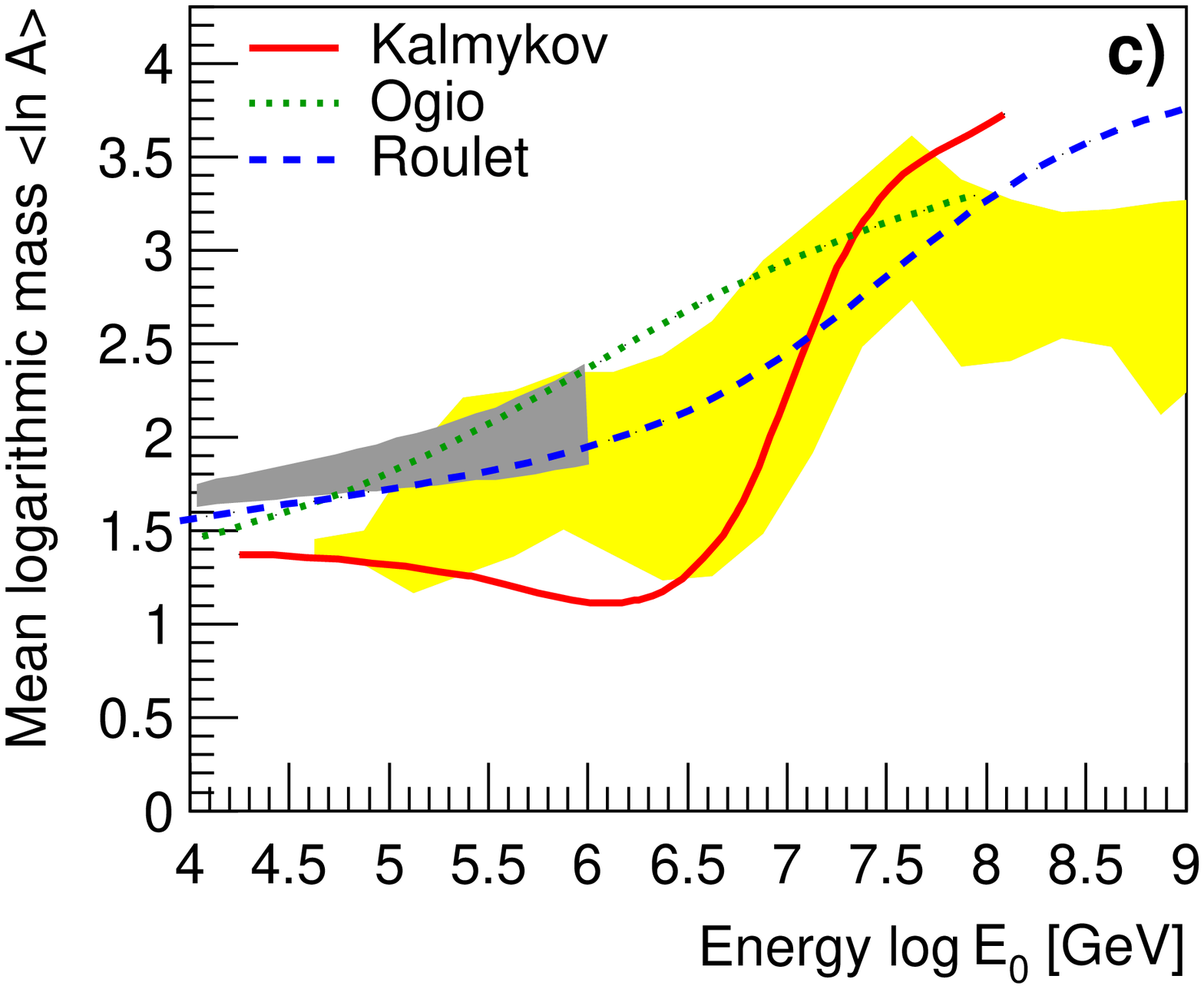,width=0.49\textwidth}
  \psfig{file=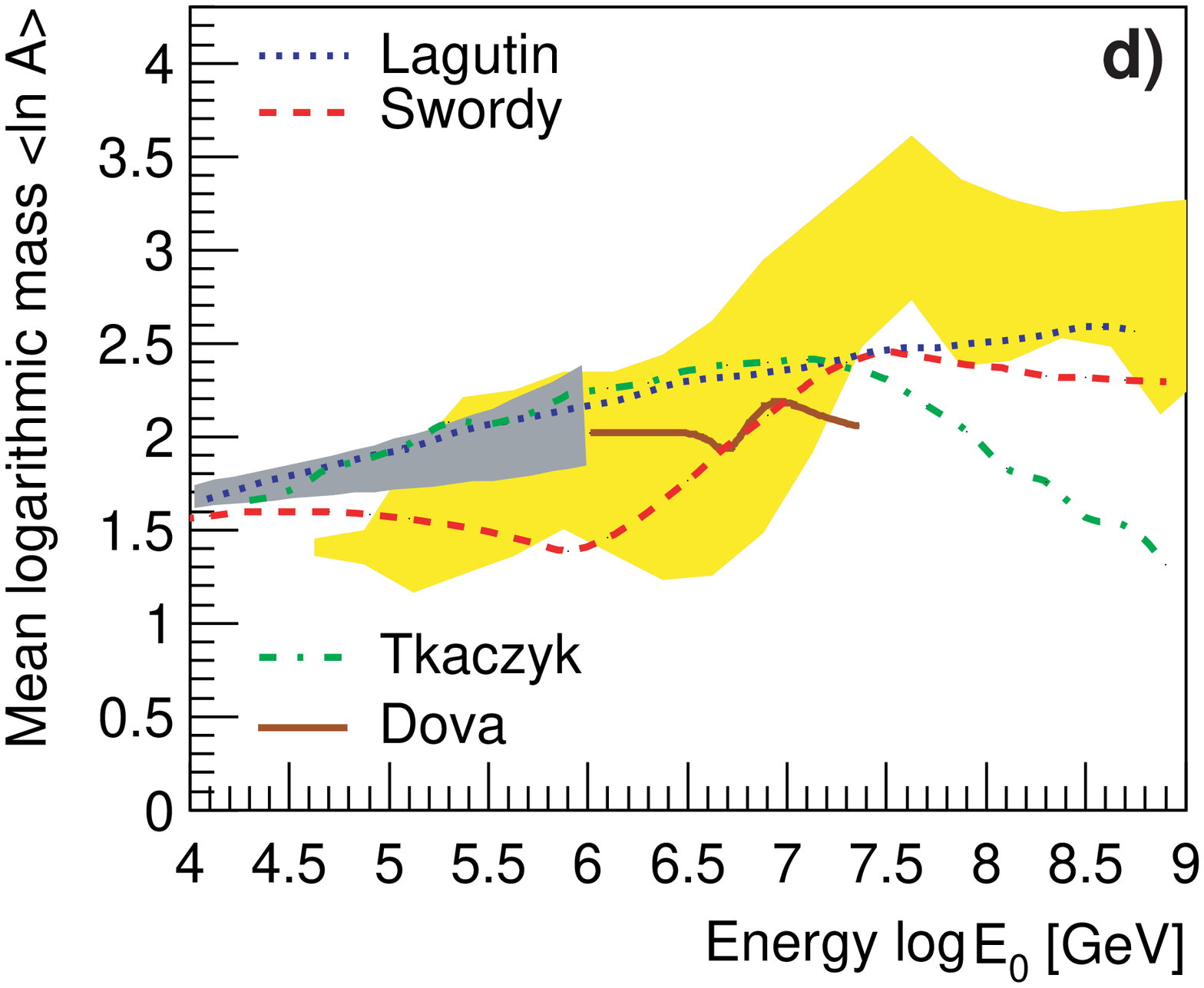,width=0.49\textwidth}
 \caption{Mean logarithmic mass as function of energy obtained by 
  direct observations (dark grey area) and air shower experiments (light grey
  area) compared with different models (lines).
  {\bf a)} Acceleration in SNRs\Cite{berezhko,kobayakawa,stanev,sveshnikova};
  {\bf b)} acceleration in GRBs\Cite{plaga,wick,dar},
           single source model\Cite{wolfendale}, 
           reacceleration in the galactic wind\Cite{voelk};
  {\bf c)} diffusion in Galaxy\Cite{kalmykov,ogio,roulet};
  {\bf d)} propagation in the Galaxy\Cite{lagutin,swordy}, as well as
	   interaction with background photons\Cite{tkaczyk} and
	   neutrinos\Cite{dova}.  
  For details see\Cite{origin}.} 
 \label{lnamod}
\end{figure}

The average experimental values from both classes of air shower measurements
presented in \fref{masse} are shown as light grey area in \fref{lnamod}.  It
represents the mean value $\pm1$ standard deviation. The dark grey area
represents the results of direct measurements above the atmosphere.  This
experimental situation will be compared to predictions of various models in the
next section.

A different interpretation of the experimental results is given in \fref{knie}
(\rright). The average of the flux values shown in the left panel is displayed
by the data points.  The spectra for elemental groups are presented according
to a parameterization of the measurements\cite{pg}, which corresponds to the
solid lines in Figs.~\ref{elementspek} and \ref{masse}, where the agreement
with the data has been discussed.  Also shown is a proposed contribution of
ultra-heavy elements ($Z>28$), extrapolated from measurements at GeV
energies.  The individual spectra exhibit a cut-off at $E_Z=Z\cdot4.5$~PeV.
The cut-off for the heaviest elements agrees with the energy of the second knee
at $\sim400$~PeV, which is interpreted as the end of the galactic CRs, while
the knee is caused by the cut-off of the light elements.  The sum spectrum of
all elements is given by the solid line, which fits nicely the average measured
spectrum up to 100~PeV.  At low energies where the nuclei traverse a large
amount of matter ($\sim10$~g/cm$^2$), heavy nuclei are more likely to interact
with the interstellar matter as compared to light elements
($\sigma_{inel}\propto A^{2/3}$) and the spectra observed at earth are expected
to be slightly flatter for heavy nuclei. At the respective knees $\Lambda$ is
less than 1~g/cm$^2$, thus for the heaviest elements around 400~PeV more than
40\% of the nuclei are expected to survive without interaction\cite{jrhprop}.

\section{THE KNEE IN THE ENERGY SPECTRUM} \label{kneesect}

The bulk of CRs is assumed to be accelerated in strong shock fronts of
SNRs\cite{blanford}. The finite lifetime of a shock front ($\sim 10^5$~a)
limits the maximum energy attainable for particles with charge $Z$ to
$E_{max}\sim Z\cdot (0.1 - 5)$~PeV.  Many versions of this scenario have been
discussed\cite{berezhko,stanev,kobayakawa,sveshnikova}. The models differ in
assumptions of properties of the SNRs like magnetic field strength, available
energy etc.  This yields differences in $\lnA$, as can be inferred from
\fref{lnamod}a.  While older models\cite{stanev} limit the maximum energy to
about 0.1~PeV, recent ideas\cite{sveshnikova}, taking into account latest
observations of SNRs, predict maximum energies above 1~PeV.  In such a model
sufficient energy is released from SNRs to explain the observed spectra, see in
\fref{elementspek} the calculations by Sveshnikova \etal\  A special case of
SNR acceleration is the single source model\cite{wolfendale}, which predicts in
the knee region pronounced structures in the all-particle energy spectrum,
caused by a single SNR. Such structures can not be seen in the compilation of
\fref{knie}. 

In the literature also other acceleration mechanisms, like the acceleration of
particles in $\gamma$-ray bursts, are discussed\cite{plaga,wick,dar}.  They
differ in their interpretation of the origin for the knee.  The approach by
Plaga, assuming Fermi acceleration in a "cannon ball" is not compatible with
the measured $\lnA$ values, see \fref{lnamod}b.  A different interpretation of
acceleration in the cannon ball model yields -- at the source -- a cut-off for
individual elements proportional to their mass due to effects of relativistic
beaming in jets.  The predictions of the actual model are compatible with
recent data\cite{dar}. However, it remains to be clarified how a detailed
consideration of the propagation processes, e.g., in a diffusion model, effects
the cut-off behavior observed at earth.  Gamma-ray bursts as a special case of
supernova explosions are proposed\cite{wick} to accelerate CRs from 0.1~PeV up
to the highest energies ($>10^{20}$~eV). In this approach the propagation of
CRs is taken into account and the knee is caused by leakage from the
Galaxy leading to a rigidity dependent cut-off behavior.

The propagation is accompanied by leakage of particles from the Galaxy. With
increasing energy it becomes more and more difficult to confine the
nuclei to the Galaxy.  As mentioned above, the pathlength decreases as
$\Lambda\propto E^{-\delta}$.  Such a decrease will ultimately lead to a
complete loss of the particles, with a rigidity dependent cut-off of the flux
for individual elements.  Many approaches have been undertaken to describe the
propagation process\cite{ptuskin,ogio,roulet,swordy,lagutin}.  The Leaky Box
model\cite{swordy} and the anomalous diffusion model\cite{lagutin} yield
cut-offs significantly weaker than the data shown in
\fref{elementspek} \cite{origin}.

The propagation as described in diffusion models\cite{kalmykov,ogio,roulet}
yields $\lnA$-values which are presented in \fref{lnamod}c.  The models are
based on the same principal idea\cite{ptuskin}, but take into account different
assumptions on details of the propagation process, like the structure of
galactic magnetic fields etc. This results in a more or less strong cut-off for
the flux at the individual knees and, accordingly, in a more or less strong
increase of $\lnA$. The model by Kalmykov \etal\cite{kalmykov} has been used to
describe the observed spectra in \fref{elementspek}.

During the propagation phase, reacceleration of particles has been suggested at
shock fronts in the galactic wind\cite{voelk}. Also this mechanism yields a
rigidity dependent cut-off.

Another hypothetical explanation for the knee are interactions of CRs with
background particles like massive neutrinos\cite{dova,wigmans} or photo
disintegration in dense photon fields\cite{tkaczyk,candia}.  Such models appear
to be excluded with a high level of confidence.  The interactions would produce
a large amount of secondary protons, which results in a light mass composition
at high energies, not observed by the experiments, see \fref{lnamod}d.
Furthermore, a massive neutrino, proposed in\cite{dova,wigmans} can be excluded
by measurements of the WMAP and 2dFGRS experiments\cite{hannestad}.

A completely different reason for the knee is the idea to transfer energy in
nucleon-nucleon interactions into particles, like gravitons\cite{kazanas} or
extremely high-energy muons\cite{petrukhin}, which are not observable (or not
yet observed) in air shower experiments. The latter proposal seems to be
excluded by recent measurements of the Baikal experiment\cite{baikal-mu}
setting upper limits for the flux of muons above $10^5$~GeV.

\section{CONCLUSION} 
During the last decade significant progress has been made in the measurement of
galactic CRs.  Summarizing the large number of experimental observations, there
are indications for a standard picture.  At least a large fraction of CRs seems
to be accelerated in supernova remnants up to energies of $Z\cdot(0.1-5)$~PeV.
Higher energies may be reached in additional sources, such as $\gamma$-ray
bursts.  The elemental composition of the accelerated material is extremely
similar to that in the solar system.  The particles propagate in a diffusive
process through the Galaxy.  With rising energy the pathlength decreases and
the particles escape easier from the Galaxy. This brings about the knee in the
energy spectrum. The general shape of the energy spectra should be determined
by the propagation process, maybe slightly modulated by properties of the
source spectra.


\section*{Acknowledgments}
The author is very much indebted to the cited and uncited colleagues for
building sophisticated instruments, extracting the data, and interpreting the
measurements -- a premise for such an overview.
It is a pleasure to thank the organizers for the invitation to participate in
an interesting and stimulating scientific symposium, which was held in a
pleasant environment.  
I would like to acknowledge fruitful scientific discussions with my colleagues
from the KASCADE-Grande and TRACER experiments.












\end{document}